\documentclass[10pt,twocolumn,letterpaper]{article}

\usepackage{cvpr} 
\usepackage{times}
\usepackage{epsfig}
\usepackage{graphicx}
\usepackage{amsmath}
\usepackage{amssymb}

\begin{document}

\title{Discrete Optimal Transport is a Strong Audio Adversarial Attack}

\author{Anton Selitskiy\\
 University of Rochester\\
Rochester, NY, USA\\
{\tt\small aselitsk@ur.rochester.edu}
\and
Shaikh Akib Shahriyar\\
Rochester Institute of Technology\\
Rochester, NY, USA\\
{\tt\small as8751@rit.edu}
\and
Jishnuraj Prakasan\\
Rochester Institute of Technology\\
Rochester, NY, USA\\
{\tt\small  jrpjishnuraj@gmail.com}
}

\maketitle
\thispagestyle{empty}

\begin{abstract}

In this paper, we investigate discrete optimal transport (DOT) as a black-box  attack against modern automatic speaker verification (ASV) and anti-spoofing countermeasure (CM) systems.

Our attack operates as a post-processing distribution-alignment step. Frame-level WavLM embeddings of generated speech (or another person speech) are aligned to an unpaired bona fide speech pool using entropic optimal transport and a top-$k$ barycentric projection, followed by neural vocoding. Unlike gradient-based attacks, the proposed method requires no access to model parameters, gradients, or training data.

Experiments on ASVspoof2019 and ASVspoof5 demonstrate that DOT attack substantially increases CM EER and substantially degrades ASV performance across multiple spoofing attacks. The attack transfers across datasets and remains effective after CM fine-tuning. Analysis using speaker similarity, Fréchet Audio Distance, and visualization of embedding distributions suggests that DOT succeeds by shifting source speech toward bona fide regions of the representation space rather than by maximizing speaker similarity.

These results indicate that optimal-transport-based distribution alignment represents a previously underexplored attack vector for contemporary ASV and anti-spoofing systems.
\end{abstract}


\section{Introduction}

Automatic speaker verification (ASV) systems are increasingly deployed in authentication, voice assistants, call centers, and remote identity verification. Their security relies not only on the speaker verification backend itself but also on spoofing countermeasures (CMs) designed to distinguish bona fide speech from synthetic or manipulated audio.

Recent advances in neural text-to-speech (TTS), voice conversion (VC), and speech generation have substantially improved the realism of synthetic speech, motivating the ASVspoof challenge which established standardized datasets and evaluation protocols (e.g., ASVspoof2019~\cite{wang2020asvspoof}, ASVspoof5~\cite{wang2024asvspoof}) for training and benchmarking spoofing countermeasure (CM) systems, and the development of modern countermeasures such as AASIST~\cite{Jung2021AASIST}. Although these systems achieve strong performance on benchmark datasets, their robustness against unseen generation methods and post-processing attacks remains an active research problem.

Prior attacks against ASV and anti-spoofing systems can be broadly divided into two categories. The first consists of optimization-based adversarial attacks that exploit model gradients, score feedback, or surrogate models such as Malafide~\cite{malafide} (e.g., $A18_5$  in Table~\ref{tab:2}, see~\cite[Sec.~7]{wang2024asvspoof}). The second relies on increasingly realistic speech synthesis systems whose outputs naturally resemble bona fide speech. In practical deployments, however, countermeasures are often accessible only as black boxes, limiting the applicability of gradient-based attacks and motivating attack strategies that do not require access to model internals.

In this paper, we investigate a different attack paradigm based on \emph{distribution alignment}. Rather than directly manipulating countermeasure outputs, we modify generated speech so that its embedding distribution becomes closer to that of bona fide speech.

Modern countermeasures are typically trained discriminatively to separate bona fide and spoofed distributions. Consequently, attacks that alter the global distribution of generated speech may reduce detection performance even without optimizing directly against the CM.

Our approach is motivated by recent advances in optimal-transport-based voice conversion, including 
SinkVC~\cite{VCOT}, MKL-VC~\cite{got}, and $k$DOT-VC~\cite{DVCOT}. While these methods were originally developed for speaker adaptation, the security implications of optimal transport for anti-spoofing systems have not been systematically studied.

The central hypothesis of this work is that moving generated speech toward the distribution of bona fide speech can substantially reduce the effectiveness of spoofing countermeasures. To test this hypothesis, we use discrete optimal transport (DOT) to align frame-level WavLM~\cite{wavlm} embeddings of generated utterances with embeddings extracted from an \emph{unpaired} pool of bona fide recordings. The transported embeddings are then reconstructed into waveform audio using a neural vocoder. The resulting method, which we refer to as the \emph{$k$DOT attack}, acts as a post-processing stage applicable to the output of arbitrary speech generation systems. Importantly, the attack requires neither access to countermeasure parameters nor knowledge of the training data used by the target system.

We consider a black-box threat model in which the adversary possesses generated speech (or non-target recording) and an unpaired collection of bona fide recordings from the target domain or a related domain.  Because the attack operates entirely through representation-level distribution alignment, it does not require gradients, score feedback, or iterative interaction with the target countermeasure.

We evaluate the proposed attack on ASVspoof2019 and ASVspoof5 using AASIST-based countermeasures. Beyond conventional CM metrics, we analyze attack effectiveness using ASV performance, speaker similarity, and Fréchet Audio Distance (FAD). Our experiments show that optimal-transport-based post-processing substantially increases attack success rates, transfers across datasets and countermeasure configurations, and remains competitive even after countermeasure fine-tuning. These results suggest that distribution-level attacks constitute an important and underexplored threat model for future anti-spoofing systems.

The main contributions of this work are as follows:

\begin{itemize}
\item We introduce a black-box adversarial attack against anti-spoofing systems based on discrete optimal transport and distribution alignment.
\item We demonstrate that optimal-transport post-processing substantially degrades the performance of modern spoofing countermeasures across ASVspoof2019 and ASVspoof5 benchmarks.
\item We analyze attack effectiveness using CM EER, ASV EER, speaker similarity, and distributional metrics, providing insight into why distribution alignment leads to successful attacks.
\item We investigate attack transferability and robustness to countermeasure fine-tuning, highlighting practical limitations of current anti-spoofing defenses.
\item We provide an empirical comparison between OT-based post-processing and existing speech-generation attacks, and discuss the role of vocoder mismatch and domain adaptation in adversarial transferability.
\end{itemize}

The code and audio samples are available at {\small \url{https://anton-selitskiy.github.io/DOTattack}}.


\section{Problem setup and threat model}
\label{sec:threat}

We consider an audio assistant or ASV pipeline equipped with a spoofing CM. Given an input waveform $x$, the CM outputs a score $f(x)$, which is compared against a threshold $\tau$ to yield a binary decision (bona fide vs.\ spoof). A separate ASV system has registered speakers and measures the attribution (similarity) of the input speech to the registered speakers.

\textbf{Adversary goal.}
Given a source utterance $x$ and a real speech sample $y$, the adversary aims to transform $x$ into an attacked sample $\hat{y}$ such that (i) $\hat{y}$ is accepted as bona fide by the CM, and (ii) intelligibility and naturalness are preserved (and, when $x$ represents a different speaker, the acoustic characteristics of $y$ are approximated). In ASV scenarios, the attack additionally seeks to maximize similarity to a target enrolled speaker so that the attacked signal can both evade the CM and be accepted by the downstream speaker-verification system.  

\textbf{Adversary knowledge and access.}
We assume a \emph{black-box} CM: the adversary has no access to model internals or gradients, and score or label access is optional and not required. The adversary may generate $x$ using any upstream TTS or VC system and possesses an \emph{unpaired} pool of bona fide speech from the target domain or a close proxy. We consider two  scenarios. In the synthetic-source setting, the attack begins from speech generated by a TTS or voice conversion system. In the real-source setting, the attack begins from a bona fide recording of a non-target speaker. In both cases, the objective is to transform the source signal so that it is accepted by the CM and attributed to the target speaker by the ASV system.  

\begin{figure}[t]
  \centering
  \includegraphics[width=\linewidth]{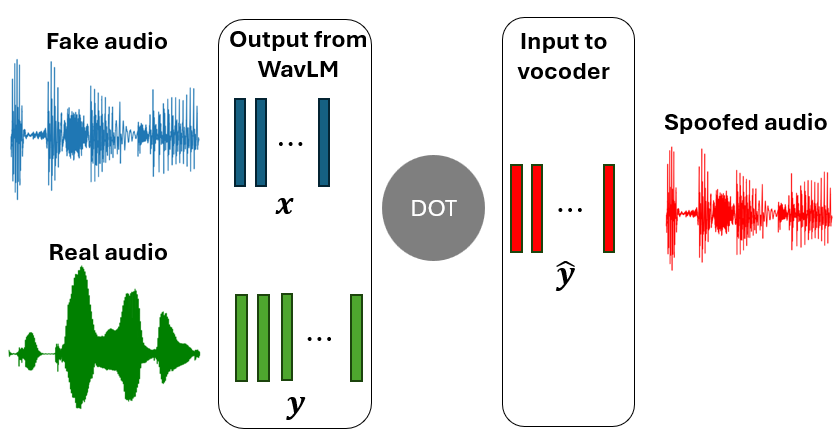}
  \caption{Schematic overview of the DOT-based voice conversion attack pipeline.}
  \label{fig:scheme}
\end{figure}

\textbf{Adversary capabilities (DOT attack).}
The adversary embeds $x$ into frame-level representations $X=\{x_i\}_{i=1}^M$ and embeds an unpaired bona fide pool $Y=\{y_j\}_{j=1}^N$. An \emph{entropic discrete OT} problem is solved to obtain a coupling matrix $\gamma \in \mathbb{R}^{M \times N}$ under the cosine cost
\begin{equation}
c(x,y) = 1 - \cos(x,y).
\end{equation}
A \emph{top-$k$ barycentric projection} is then applied to produce transported embeddings $\hat{Y}=\{\hat{y}_i\}_{i=1}^M$, which are converted back to a waveform $\hat{y}$ using a neural vocoder. The pipeline relies solely on generic pretrained components and unpaired speech data (see Fig.~\ref{fig:scheme}).

\textbf{Perceptual constraints.}
The attack aims to preserve the linguistic content of $x$ while introducing acoustic characteristics derived from the bona fide speech pool $y$, increasing similarity to the target speaker, and achieving CM evasion. We do not impose explicit perturbation norms (e.g., $\ell_p$ constraints), as the proposed method operates via \emph{distributional alignment} rather than pointwise perturbations.

We follow the ASVspoof evaluation protocols~\cite{wang2020asvspoof,wang2024asvspoof}. The effectiveness of the proposed attack is assessed from both the countermeasure (CM) and automatic speaker verification (ASV) perspectives. For spoofing detection, we report the CM equal error rate (EER), tandem detection cost function (t-DCF), and minimum t-DCF (min t-DCF), which quantify the degradation of the CM and the overall impact on the tandem ASV-CM system. To evaluate the ability of the attack to impersonate a target speaker, we report the ASV equal error rate (ASV EER), which measures verification performance under attack, together with a speaker similarity metric (SIM), which quantifies preservation of target speaker characteristics. In addition, we study attack transferability across datasets and robustness to CM fine-tuning. We further analyze practical factors affecting attack performance, including vocoder overlap between training and evaluation data. The influence of the top-$k$ barycentric projection parameter $k$ and target utterance duration has been investigated in~\cite{DVCOT}. 

\section{Methodology: Discrete OT Attack}
\label{sec:format}

\subsection{Discrete Optimal Transport-Based Voice Conversion}

Recent voice conversion methods operate on frame-level embeddings extracted by pretrained speech representations such as WavLM~\cite{wavlm}.  Let
$
\mathbf{x} = [x_1, x_2, \ldots, x_M]
$
denote the sequence of embeddings extracted from a source speaker recording, where $x_i \in \mathbb{R}^{1024}$ are obtained using the pretrained WavLM Large model. WavLM encodes audio using a window length of $25$~ms and a hop size of $20$~ms. Similarly, let
$
\mathbf{y} = [y_1, y_2, \ldots, y_N]
$
denote the embeddings of a target speaker utterance. 

The first embedding-based approach, $k$NN-VC~\cite{VCNN}, converts each source embedding $x_i$ to $\hat{y}_i$ by averaging its nearest neighbors among $\{y_j\}_{j=1}^{N}$ under cosine similarity. Subsequent work~\cite{VCOT} replaced nearest-neighbor weights with transport probabilities  obtained from discrete optimal transport (OT), resulting in the method called \textit{SinkVC}  (from the Sinkhorn algorithm commonly used to solve DOT problem). 

More recently,~\cite{got} proposed approximating OT by a Gaussian transport map, leading to computationally efficient voice conversion methods based on continuous OT. The authors refer to this method as \emph{factorized MKL} (Monge -- Kantorovich Linear). However, these approximations typically sacrifice conversion quality and speaker similarity.

In this work, we employ the recently proposed $k$DOT-VC method~\cite{DVCOT}, which constructs a transport map using a truncated barycentric projection of discrete OT. Previous work showed that $k$DOT-VC better approximates the target speaker distribution than existing WavLM-based VC methods while preserving speech intelligibility. 

\subsection{Discrete OT and barycentric projection}
\label{subsec:2.1}
Assume  there are $M$ vectors in $X$ and $N$ vectors in $Y$ with probability masses $p_i=\mathbf{P}(x_i)$ and $q_j=\mathbf{Q}(y_j).$ The joint distribution $\pi(x,y)$ is represented as a non-negative matrix $\gamma$ with $\gamma_{ij}=\pi(x_i,y_j),$ $i=1,\ldots,M$ and $j=1,\ldots,N.$
The goal of optimal transport (OT) is to find the joint distribution $\pi$ known as \textit{Kantorovich plan,}   that minimizes the expected transport cost
\begin{equation}\label{eq:dot}
    \sum_{i}^{M} \sum_{j}^{N} \gamma_{ij} c(x_i,y_j) \to \underset{\gamma_{ij}}{\min},
\end{equation}
subject to the marginal constraints:
\begin{equation}
    p_i = \sum_{j=1}^{N} \gamma_{ij}\quad \text{and}\quad q_j = \sum_{i=1}^{M} \gamma_{ij}.
\end{equation}
Given a solution $\gamma$, a transport map can be defined via the \textit{barycentric projection} 
\begin{equation}
\hat{y}_i = \tilde{\gamma}_{i1} y_1 + \tilde{\gamma}_{i2} y_2 +\ldots +\tilde{\gamma}_{iN} y_N.  
\end{equation} 
where $\tilde{\gamma}_{ij} = \frac{\gamma_{ij}}{p_i}.$
This transform can be interpreted as the conditional expectation $\mathbf{E}[y|x=x_i].$

Since the underlying distributions of speaker embeddings are unknown, we  use  \textit{empirical distributions:} 
$\mathbf{P}(x_i)=\dfrac{1}{M}$ and  $\mathbf{P}(y_j)=\dfrac{1}{N}.$ For each $x_i,$ we sort the target embeddings $y_j$ in decreasing order of $\gamma_{ij},$  denoting  the sorted vectors as $y^{sort(i)}_j.$ The sorted coupling weights along each row (with fixed $i$) are denoted  by $\gamma^{sort}_{ij}.$
We define the $k$DOT mapping as the baricentric projection of the OT map over top-$k$ vectors,
\begin{equation}
    x_i \overset{T}{\mapsto} \hat{y}_i  = \sum_{j=1}^{k} \tilde{\gamma}^{sort}_{ij}y^{sort(i)}_j,\quad \tilde{\gamma}^{sort}_{ij} = \frac{\gamma^{sort}_{ij}}{\sum_{s=1}^{k}\gamma^{sort}_{is}}.
\end{equation}


By retaining only the $k$ largest transport normalized weights $\tilde{\gamma}_{ij}$, we obtain  $k$DOT-VC method. We use $k=5$; as~\cite{DVCOT} shows, the conversion quality remains stable for $3 \leq k \leq 10$ and degrades for larger values of $k$.

The intuition behind this attack is simple: the approximate map $T$ shifts the empirical distribution of generated speech toward the target (real speech) distribution. Since CMs are trained to reject synthetic distributions and accept real speech, moving the generated distribution closer to the real one can reduce detection performance, i.e., produce a strong, dataset- and model-agnostic adversarial effect.



\subsection{Attack Variants}

For the synthetic-source attack scenario, the source set $X$ consists of generated utterances from the development partition of ASVspoof2019. To construct the bona fide target pool $Y$, we investigate two alternatives corresponding to different recording conditions and speaker populations.

\textbf{OT$_1$.}
The target pool $Y$ is formed from bona fide recordings in the ASVspoof2019 dataset, which are based on the VCTK corpus~\cite{vctk}. This setting represents a matched-domain attack in which source and target speech originate from similar recording conditions.

\textbf{OT$_2$.}
The target pool $Y$ is constructed from the LibriSpeech train-clean-100 corpus~\cite{panayotov2015librispeech}. This choice is motivated by the fact that ASVspoof5 is based on LibriVox recordings, from which LibriSpeech is derived. We select the first 40 speakers (ordered by speaker ID) and randomly sample 10 utterances per speaker. The selected utterances are sorted by duration and concatenated to increase the amount of target speech available for transport.

Previous work has shown that voice conversion quality improves with increasing target-speaker duration~\cite{DVCOT}. The LibriSpeech-based target pool therefore enables the construction of longer speaker representations through utterance concatenation. As shown in Table~\ref{tab:1}, the performance difference between OT$_1$ and OT$_2$ is relatively small. Consequently, we adopt OT$_2$ as the default attack configuration until Sec.~\ref{sec:5.4}, where we evaluate ASVspoof2019.  

\textbf{OT$_{19}$.} In Sec.~\ref{sec:5.4}, we apply $k$DOT-VC in its original voice conversion setting, converting bona fide speech from non-target speakers to target speakers on ASVspoof2019. Despite not being designed as an adversarial method, this conversion pipeline proves highly effective at bypassing the countermeasure.

\section{Experimental Setup}
\label{sec:6}

\textbf{Datasets.}
We use three public corpora. For construction of the bona fide target pool used by the DOT attack (OT$_2$), we draw recordings from the LibriSpeech train-clean-100 subset~\cite{panayotov2015librispeech,kaggleLS}. Countermeasure evaluation and cross-dataset transfer experiments are conducted on ASVspoof2019~\cite{wang2020asvspoof,kaggleAS} and ASVspoof5~\cite{wang2024asvspoof}. Unless otherwise specified, we use the official train, development, and evaluation partitions provided with each benchmark.

\textbf{Embeddings.}
Frame-level speech representations are extracted using WavLM Large~\cite{wavlm}. Following the $k$NN-VC framework~\cite{VCOT}, we use embeddings from the sixth transformer layer.

\textbf{Optimal Transport.}
Discrete optimal transport is solved using entropic regularization and the Sinkhorn algorithm implemented in the \texttt{POT} library~\cite{pot}. Unless otherwise stated, the default regularization parameter is used.

\textbf{Vocoder.}
Transported embeddings are converted back to waveform audio using HiFi-GAN, employing the implementation distributed with the $k$NN-VC framework~\cite{VCOT}.

\textbf{Countermeasure.}
We evaluate attack effectiveness using the official AASIST implementation~\cite{Jung2021AASIST} and its pretrained variants.

\textbf{Metric 1: Countermeasure Performance.}
We report the equal error rate (CM EER), tandem detection cost function (t-DCF), and minimum tandem detection cost function (min t-DCF), which are the standard metrics used in the ASVspoof evaluations~\cite{wang2020asvspoof,wang2024asvspoof}.

\textbf{Metric 2: Distributional Similarity.}
To quantify distributional changes induced by DOT independently of a specific countermeasure, we compute the Fréchet Audio Distance (FAD)~\cite{fadtk}. VGGish embeddings are extracted using \texttt{torchvggish} (v0.2)~\cite{vgg}; implementation details follow~\cite{DVCOT}.


\textbf{Metric 3: Speaker Similarity (SIM).}
Speaker similarity is measured as the cosine similarity between enrollment and test utterance embeddings extracted using an ECAPA-TDNN~\cite{desplanques2020ecapa} speaker encoder. Higher values indicate stronger similarity to the target speaker.

\textbf{Metric 4: Speaker Verification Performance.}
To evaluate the ability of the attack to impersonate a target speaker, we report ASV EER using an ECAPA-TDNN speaker verification backend. This metric complements SIM by measuring verification performance under attack conditions.

\section{Analysis and Evaluation}
\label{sec:pagestyle}

For evaluation, we used the AASIST model~\cite{Jung2021AASIST}, pretrained on the ASVspoof2019 dataset (denoted as AASIST$_{2019}$), and on ASVspoof5 (denoted as AASIST$_5$). 

Table~\ref{tab:1} reports equal error rates (EER) for the generation algorithm A18 from ASVspoof2019 and attack A18 in ASVspoof5, which we denote by A18$_5$, as well as for the two attacks introduced in the previous section. Algorithms A18 and A18$_5$ were selected because they exhibit the highest EER within their respective datasets.

\begin{table}[h]
 \begin{center}
 \setlength{\tabcolsep}{3pt}

\begin{tabular}{|lcccc|} 
    \hline
    Attack & AASIST$_{2019}$  & AASIST$_{2019}^{FT}$ & AASIST$_5$   & AASIST$_5^{FT}$ \\
    \hline
    \hline
    A18 &  2.614 & 3.141 & 77.735 & 44.951 \\
    A18$_5$ & 0.435 & 1.443 & 57.933 & 2.730\\
    OT$_1$  &  11.111   &  -  & 7.268 & -  \\
    OT$_2$    & 7.925 & 0.216 &  11.180 & 12.586  \\
    \hline
\end{tabular}
\end{center}
\caption{EER$\downarrow$ for strongest attacks  and  proposed attacks.}
\label{tab:1}
\end{table}

Column AASIST$_{2019}$ shows the EER on the validation subset of ASVspoof2019, evaluated with AASIST$_{2019}$ (i.e., using bonafide data from the ASVspoof2019 validation set).

Column AASIST$_5$ presents the EER computed with AASIST$_5$. A notable observation emerges: the strongest attack proposed in ASVspoof5 is detected robustly by the model pretrained on ASVspoof2019. Conversely, most methods from ASVspoof2019 achieve higher EERs than the novel attacks in ASVspoof5 (see also columns (2019) and (5) in Table~\ref{tab:2}). This asymmetry reflects \emph{adversarial transferability} across models/datasets~\cite{szegedy2014intriguing,goodfellow2015explaining,papernot2016transferability}, observed previously in speech systems as well~\cite{alzantot2018did, asv}.

Column AASIST$_{2019}^{FT}$ shows results after fine-tuning AASIST$_{2019}$ by including OT$_2$ examples in the training set of ASVspoof2019.  

The final column reports results after fine-tuning AASIST$_5$ using OT$_2$ and A18$_5$ data (see Sec.~\ref{sec:ft} for details).

Table~\ref{tab:2} extends this analysis by including attacks A07--A19 (ASVspoof2019) and A17$_5$--A31$_5$ (ASVspoof5). Columns (2019) and (5) represent baseline evaluation with AASIST$_{2019}$ and AASIST$_5$, while columns (2019$_{OT}$) and (5$_{OT}$) show results after applying optimal transport to A07--A19 data.

\subsection{Fine-tuning}
\label{sec:ft}

Since all generation methods and attacks from ASVspoof5 resulted in very low EER when detected by AASIST$_{2019}$, we fine-tuned AASIST$_{2019}$ only with OT$_2$ data. Examples were obtained by applying optimal transport to generated audio in the ASVspoof2019 training set.

Comparing columns (2019$_{OT}$) and (2019$_{OT}^{FT}$) in Table~\ref{tab:2} (equivalently, columns AASIST$_{2019}$ and AASIST$_{2019}^{FT}$ in Table~\ref{tab:1}), we see that OT$_2$ attacks are easily detected after fine-tuning. The EER for A18$_5$ remained largely unchanged.

For AASIST$_5$, which suffers from a strong A18 attack (and also from A19, though here we focus on adversarial attacks), we fine-tuned using a subset of A18$_5$ data from the ASVspoof5 evaluation set (12,000 of 27,000 recordings to keep the proportion of classes, with the remainder reserved for evaluation) and included OT$_2$ training data used for fine-tuning AASIST$_{2019}$.  

The comparison between AASIST$_5$ and AASIST$_5^{FT}$ (Table~\ref{tab:1}) shows that A18$_5$ becomes well detected after fine-tuning. However, the OT$_2$ attack still maintains a high EER.

\subsection{The role of the vocoder}
The relatively low EER for OT$_2$ in AASIST$_5$ (Table~\ref{tab:1}) compared with other attacks (see also Table~\ref{tab:2}, column (5), rows A17$_5$--A31$_5$) may be explained by vocoder overlap. Most of the ASVspoof5 training data are generated with the HiFi-GAN vocoder, which is also used in our attack. This likely explains the elevated EER for methods employing different vocoders, particularly the extremely high EER observed for ASVspoof2019 methods evaluated with AASIST$_5$ (Table~\ref{tab:2}, column (5), rows A07--A19).

\begin{figure}[t]

  \centering
  \includegraphics[width=8cm]{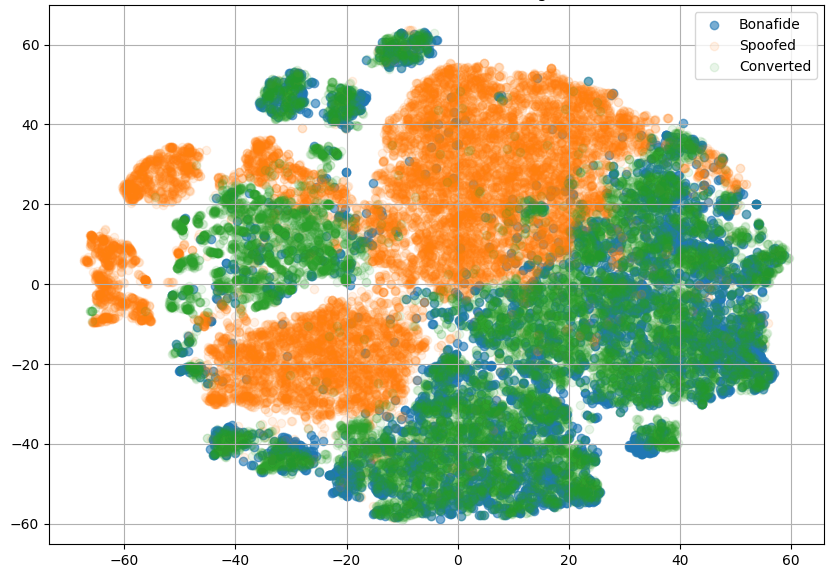}
\caption{Bona fide embeddings from LibriSpeech.}
\label{fig:res}
\end{figure}

\begin{figure}[t]

  \centering
  \includegraphics[width=8cm]{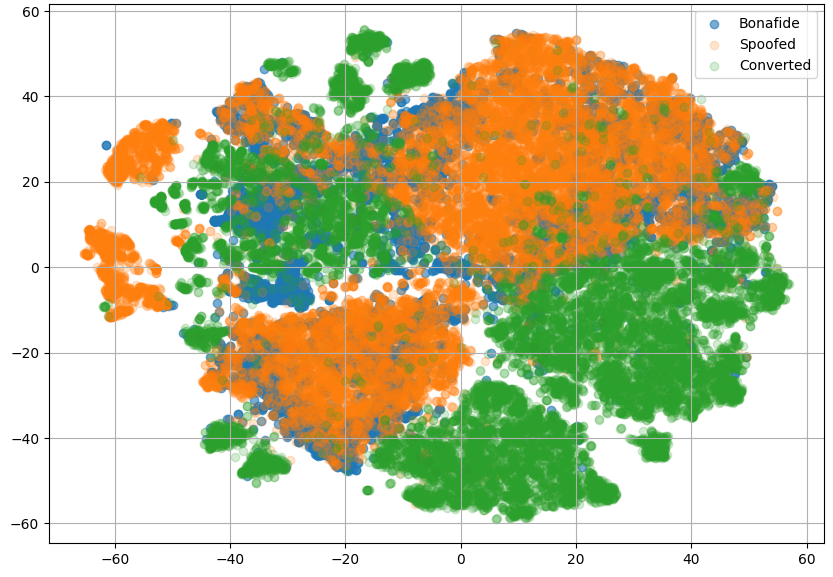}
\caption{Bona fide embeddings  from ASVspoof2019.}
\label{fig:resTR}
\end{figure}

\begin{table}[h]
\begin{center}
\small
\setlength{\tabcolsep}{4pt}
\begin{tabular}{|lcccccc|} 
    \hline
    Attack & 2019 & $2019_{OT}$ & 2019$^{FT}_{OT}$ & 5 & $5_{OT}$  & 5$^{FT}_{OT}$ \\
    \hline
    \hline
    A07  &  0.52  &   0.51 & 0.12  & 33.28 &  4.39 &  10.05\\
    A08   &    0.42 & 3.70 & 0.11 & 19.90 &  4.23 & 9.49\\
    A09   &    0.00 & 0.69 & 0.05 & 7.79 &  2.92 & 7.44\\
    A10   &    0.86 & 0.73 & 0.13 & 39.42 &  6.39 & 14.30\\
    A11   &    0.18 & 0.71 & 0.13 & 5.65 &  5.40 & 13.71\\
    A12   &    0.78 & 0.73 & 0.09 & 55.41 &  7.40 & 13.04\\
    A13   &    0.15 & 0.51 & 0.06 & 65.51 &  6.38 & 15.07\\
    A14   &    0.15 & 0.81 & 0.05 & 15.34 &  4.34 & 8.15\\
    A15   &    0.55 & 0.77 & 0.05 & 7.30 &  3.94 & 8.88\\
    A16   &    0.65 & 1.98 & 0.11 & 71.97 &  11.37 & 17.15\\
    A17   &    1.26 & 13.34 & 0.37 & 73.26 &  17.69 & 12.36\\
    A18   &    2.61 & 12.26 & 0.27 & 77.73 &  24.19 & 14.74\\
    A19   &    0.65 & 12.98 & 0.27 & 72.57 &  19.02 & 14.07\\
    A17$_5$  &  1.36  &   - &  - & 11.43 &  - &  -\\
    A18$_5$   &    0.43 & - & - & 51.92 &  - & -\\
    A19$_5$   &    0.18 & - & - & 57.93 &  - & -\\
    A20$_5$   &    0.11 & - & - & 49.78 & - & -\\
    A21$_5$   &    0.72 & - & - & 13.28 &  - & - \\
    A22$_5$   &    1.24 &  - & - & 14.07 &  - & -\\
    A23$_5$   &    0.15 & - & - & 28.81 &  - & -\\
    A24$_5$   &    1.39 & - & - & 10.69 &  - & -\\
    A25$_5$   &    0.35 & - & - & 22.29 &  - & -\\
    A26$_5$   &    1.86 & - & - & 27.41 &  - & -\\
    A27$_5$   &    0.18 & - & - & 24.10 &  - & -\\
    A28$_5$   &    0.94 & - & - & 23.57 &  - & -\\
    A29$_5$   &    0.58 & - & - & 6.83 &  - & -\\
    A30$_5$   &    0.25 & - & - & 39.89 &  - & -\\
    A31$_5$   &    0.18 & - & - & 26.53 &  - & -\\
    \hline
\end{tabular}
\end{center}
\caption{EER$\downarrow$ before finetuning (2019 and 5); after OT$_2$ attack (2019$_{OT}$ and 5$_{OT}$); after attack and finetuning (2019$_{OT^{FT}}$ and 5$_{OT^{FT}}$).}
\label{tab:2}
\end{table}

\subsection{The optimal transport property}
To further illustrate the effect of optimal transport, we applied t-SNE to VGGish embeddings.  

Figure~\ref{fig:res} shows embeddings from LibriSpeech bonafide data, ASVspoof2019 generated data (A01--A06), and their OT-transformed versions. The transformed embeddings (Converted) align closely with the LibriSpeech target distribution (Bonafide).

Figure~\ref{fig:resTR} shows embeddings from ASVspoof2019 bonafide data, ASVspoof2019 generated data, and their OT-transformed counterparts. Here, the discrepancy between ASVspoof2019 bonafide data and LibriSpeech becomes evident. This explains why the Fréchet Audio Distance (FAD) between bonafide training data and OT$_2$ is relatively large (Table~\ref{tab:fad}), and why AASIST$_{2019}$ readily detects OT$_2$ after fine-tuning, despite its high pre-finetuning EER.

\begin{table}[t]
\begin{center}
\small
\begin{tabular}{|lccc|}
\hline
BF Dataset        & BF--Spoof & BF--OT & Spoof--OT \\
\hline\hline
LibriSpeech    & 4.742     & 0.508  & 3.665     \\
ASVspoof2019   & 1.289     & 3.402  & 3.665     \\
\hline
\end{tabular}
\end{center}
\caption{FAD$\downarrow$ between (Bona fide, Spoof), (Bona fide, OT), and (Spoof, OT).}
\label{tab:fad}
\end{table}

\subsection{$k$DOT attack performance on ASVspoof2019}
\label{sec:5.4}

\subsubsection{Conversion of Real Speech}
First, we evaluate the proposed OT-based attack from real speech on the ASVspoof2019 logical access benchmark. For each target speaker, we exclude seven utterances from the evaluation partition and use them as the target speaker reference set for the OT attack. Importantly, enrollment recordings are not used during attack construction, ensuring that the enrollment data remain unseen and preventing trivial leakage between attack generation and ASV evaluation.

Table~\ref{tb:before} reports the baseline performance of the original ASVspoof2019 attack types together with the standalone OT$_{19}$ attack. Several observations are immediate.

\paragraph{Countermeasure vulnerability}
The proposed OT attack achieves a CM EER of 29.67\%, which is substantially higher than all baseline attacks. In comparison, the strongest original attacks (A17--A19) achieve CM EER values between 0.65\% and 2.58\%. This indicates that $k$DOT-VC dramatically increases the probability that spoofed speech is classified as bona fide by the countermeasure.

\paragraph{ASV behavior}
The standalone OT attack yields an ASV EER of 21.19\%, indicating a substantial degradation of speaker-verification performance. However, this degradation is not solely explained by speaker similarity, as discussed below.

\paragraph{Speaker similarity}
The OT attack achieves a moderate average speaker similarity score of \(0.56 \pm 0.10\). Notably, some baseline attacks such as A16 exhibit higher similarity scores (\(0.73 \pm 0.10\)) while remaining relatively ineffective against the CM. This suggests that high speaker similarity alone is insufficient to explain countermeasure failure.

\begin{table}
\begin{center}
\small
\begin{tabular}{|l|ccc|}
\hline
Attack & CM EER & ASV EER & SIM \\
\hline\hline
A07 & $0.55$ & $28.85$ & $0.61\pm 0.09$\\
A08 & $0.43$ & $14.88$ & $0.52\pm 0.10$\\
A09 & $0.00$ & $0.91$ & $0.19\pm 0.10$\\
A10 & $0.85$ & $47.57$ & $0.69\pm 0.12$\\
A11 & $0.18$ & $44.93$ & $0.67\pm 0.11$\\
A12 & $0.73$ & $26.49$ & $0.57\pm 0.13$\\
A13 & $0.16$ & $11.85$ & $0.47\pm 0.11$\\
A14 & $0.17$ & $29.66$ & $0.62\pm 0.10$\\
A15 & $0.57$ & $37.11$ & $0.66\pm 0.09$\\
A16 & $0.65$ & $56.10$ & $0.73\pm 0.10$\\
A17 & $1.26$ & $1.09$ & $0.16\pm 0.12$\\
A18 & $2.58$ & $1.57$ & $0.24\pm 0.11$\\
A19 & $0.65$ & $3.70$ & $0.31\pm 0.12$\\
OT$_{19}$ & $29.67$ & $21.19$ & $0.56\pm 0.10$\\
\hline
\end{tabular}
\end{center}
\caption{Results on ASVspoof2019 extended with OT$_{19}$ attack.}
\label{tb:before}
\end{table}

\subsubsection{Conversion of Existing Attacks}

We next apply the OT transformation to all spoofing attacks in ASVspoof2019 by transporting each attack toward the corresponding target-speaker reference set. The resulting performance changes are summarized in Table~\ref{tb:after}.

\begin{table}
\begin{center}
\setlength{\tabcolsep}{4pt}
\small
\begin{tabular}{|l|cc|cc|cc|}
\hline
\scriptsize{Attack} &
\scriptsize{CM EER} & $\Delta$ &
\scriptsize{ASV EER} & $\Delta$ &
SIM$^{*}$ & $\Delta$ \\
\hline\hline
A07 & $0.47$ & $-0.08$ & $27.03$ & $-1.82$ & $0.60$ & $-0.01$\\
A08 & $4.94$ & $+4.51$ & $27.39$ & $+12.51$ & $0.60$ & $+0.08$\\
A09 & $0.63$ & $+0.63$ & $25.11$ & $+24.20$ & $0.59$ & $+0.40$\\
A10 & $0.59$ & $-0.26$ & $27.27$ & $-20.30$ & $0.60$ & $-0.09$\\
A11 & $0.63$ & $+0.45$ & $26.99$ & $-17.94$ & $0.60$ & $-0.07$\\
A12 & $0.71$ & $-0.02$ & $28.27$ & $+1.78$ & $0.60$ & $+0.03$\\
A13 & $0.44$ & $+0.28$ & $22.85$ & $+11.00$ & $0.57$ & $+0.10$\\
A14 & $0.75$ & $+0.58$ & $28.86$ & $-0.80$ & $0.61$ & $-0.01$\\
A15 & $0.73$ & $+0.16$ & $28.93$ & $-8.18$ & $0.61$ & $-0.05$\\
A16 & $1.94$ & $+1.29$ & $30.06$ & $-26.04$ & $0.61$ & $-0.12$\\
A17 & $26.13$ & $+24.87$ & $17.83$ & $+16.74$ & $0.54$ & $+0.38$\\
A18 & $25.46$ & $+22.88$ & $21.45$ & $+19.88$ & $0.56$ & $+0.32$\\
A19 & $24.93$ & $+24.28$ & $22.16$ & $+18.46$ & $0.57$ & $+0.26$\\
OT$_{19}$  & $29.66$ & $-0.01$ & $21.19$ & $0.00$ & $0.56$ & $0.00$\\
\hline
\end{tabular}
\end{center}
\caption{Results after applying the OT transformation to existing attacks. $\Delta$ denotes the difference relative to the original model (After -- Before). Positive CM EER and ASV EER deltas indicate a stronger attack. $^{*}$All standart deviations are between $0.09$ and $0.10$}
\label{tb:after}
\end{table}

\paragraph{Large gains for difficult attacks} 
The most striking improvements occur for attacks A17--A19. For example, A17 increases from a CM EER of 1.26\% to 26.13\%, while A18 increases from 2.58\% to 25.46\%. Similar improvements are observed in ASV EER and SIM. These results indicate that OT can substantially amplify attacks that were originally ineffective against the CM.

\paragraph{Moderate or negative changes for some attacks} 
For several attacks (e.g., A07, A10, A11), the OT transformation slightly decreases CM EER or ASV EER. This suggests that OT is not uniformly beneficial across all synthesis methods and that the interaction between the source attack distribution and the target bona fide distribution plays an important role.

\paragraph{Consistent increase in similarity}
Most attacks experience an increase in SIM after OT processing, especially A09, A17, A18, and A19. Nevertheless, the relationship between SIM and CM EER is clearly nonlinear: some attacks obtain large CM gains with only moderate similarity increases.

\subsubsection{Joint system-level impact}

To quantify the overall effect on the tandem ASV-CM system, Table~\ref{tb:joint} reports aggregate metrics before and after applying OT. The attack increases CM EER from 11.17\% to 16.41\% and nearly doubles the min-tDCF from 0.20 to 0.39, indicating a substantial degradation of system-level security. At the same time, ASV EER rises from 0.79\% to 14.88\%, confirming that the attack also disrupts speaker verification performance.
Interestingly, the average SIM increases only moderately from \(0.41 \pm 0.25\) to \(0.47 \pm 0.22\). This again suggests that the success of the attack cannot be explained solely by improved speaker imitation.
\begin{table}[h]
\begin{center}
\small
\begin{tabular}{|l|cccc|}
\hline
 & CM EER & min-tDCF & ASV EER  & SIM  \\
\hline\hline
Before & $11.17$ & $0.20$ & $0.79$  & $0.41\pm 0.25$\\
After & $16.41$ & $0.39$ & $14.88$  & $0.47\pm 0.22$\\
\hline
\end{tabular}
\end{center}
\caption{Joint ASV-CM metrics.}
\label{tb:joint}
\end{table}

\section{Conclusion} 
Taken together, the results support a consistent interpretation of the proposed attack. OT-based attacks substantially increase CM EER across multiple datasets and attack families. The effectiveness of OT is not primarily driven by maximizing speaker similarity or by exploiting artifacts of a specific synthesis model. Instead, OT acts as a distribution-alignment mechanism that shifts spoofed speech toward regions of the embedding space occupied by bona fide speech. This interpretation explains:
the substantial increase in CM EER,
the strong gains for previously difficult attacks,
the partial transferability across attack types and datasets,
and the observation that moderate similarity increases can still produce large countermeasure failures.

In other words, the OT transformation appears to improve the global statistical resemblance of spoofed speech to bona fide speech, making the resulting signals harder for the CM to distinguish even when the target speaker similarity is not maximized. Distributional alignment represents a previously underexplored vulnerability of anti-spoofing systems.

 This explains the observed transferability across datasets, the substantial increase in CM EER, and the partial robustness of the attack to countermeasure fine-tuning.

\textbf{Limitations.} The attack does not explicitly optimize speaker similarity and therefore should be viewed primarily as a countermeasure-evasion attack.

\section{Acknowledgment}
\label{sec:print}
The first author thanks You (Neil) Zhang for pointing out the adversarial attacks in ASVspoof5 dataset~\cite{wang2024asvspoof}. 
The first author also thanks Arip Asadulaev and Rostislav Korst for clarifications regarding their work~\cite{VCOT}.



{\small
\bibliographystyle{ieee}
\bibliography{mybib}

@article{wavlm,
  author={S. Chen and Ch. Wang  and  Zh. Chen and others},
  journal={IEEE Journal of Selected Topics in Signal Processing}, 
  title={{WavLM}: Large-Scale Self-Supervised Pre-Training for Full Stack Speech Processing}, 
  year={2022},
  volume={16},
  number={6},
  pages={1505-1518},}

@inproceedings{VCNN,
  author = 	 "M. Baas and  B. van Niekerk  and  H. Kamper",
  title   =      "Voice Conversion With Just Nearest Neighbors",
  booktitle =    "Interspeech",
  month = apr,
  year = 	 2023,
  volume = 	 "II",
  pages = 	 "803-806"
}

@misc{VCOT,
      title={Optimal Transport Maps are Good Voice Converters}, 
      author={Asadulaev, A. and Korst, R. and others},
      year={2024},
      note = {arXiv 2411.02402 {\tt tr.pdf}},
      eprint={2411.02402},
      archivePrefix={arXiv},
      primaryClass={cs.SD},
      url={https://arxiv.org/abs/2411.02402}, 
}

@inproceedings{
got,
title={Training-Free Voice Conversion with Factorized Optimal Transport},
author={Alexander Lobashev and Assel Yermekova and Maria Larchenko},
booktitle={Interspeech},
year={2024},
pages={1373-1377}
}

@inproceedings{DVCOT,
  author = 	 "Selitskiy, A. and Kocharekar, M.",
  title   =      "Discrete Optimal Transport and Voice Conversion",
  booktitle =    "{2026 11th International Conference on Machine Learning Technologies (ICMLT),} Berlin, Germany",
  month = may,
  year = 	 2026,
  pages = 	 ""
}

@inproceedings{malafide,
  author = 	 "Panariello, M. and Ge, W. and Tak, H. and others",
  title   =      "Malafide: a Novel Adversarial Convolutive Noise Attack Against Deepfake and Spoofing Detection Systems",
  booktitle =    "Interspeech",
  month = apr,
  year = 	 2023,
  pages = 	 "2868--2872"
}

@article{wang2024asvspoof,
  title= "{ASVspoof} 5: Design, Collection and Validation of Resources for Spoofing, Deepfake, and Adversarial Attack Detection Using Crowdsourced Speech",
  author="Wang, X. and  Delgado, H.  and  others",
  journal= "Computer Speech \& Language",
  volume= "95",
  pages="1--27",
  year= "2026",
  publisher="Elsevier"
}

@article{wang2020asvspoof,
  title= "{ASVspoof} 2019: A Large-Scale Public Database of Synthesized, Converted and Replayed Speech",
  author="Wang, Xin and Yamagishi, Junichi and  others",
  journal= "Computer Speech \& Language",
  volume= "64",
  pages="101--114",
  year= "2020",
  publisher="Elsevier"
}

@misc{Jung2021AASIST,
      title={{AASIST}: Audio Anti-Spoofing using Integrated Spectro-Temporal Graph Attention Networks}, 
      author={Jung, J. and Heo, H.-S. and others},
      year={2021},
      note = {arXiv 2110.01200},
      eprint={2110.01200},
      archivePrefix={arXiv},
      primaryClass={eess.AS},
      url={https://arxiv.org/abs/2110.01200}, 
}

@incollection{vctk,
author = {Veaux, Christophe and Yamagishi, Junichi and MacDonald, Kirsten},
title	= "{CSTR VCTK Corpus:} English Multi-Speaker Corpus for {CSTR} Coice Cloning Toolkit (version 0.92)",
year = {2017},
booktitle="",
publisher={University of Edinburgh. The Centre for Speech Technology Research (CSTR)}
}

@inproceedings{panayotov2015librispeech,
  title={{LibriSpeech}: An {ASR} Corpus Based on Public Domain Audio Books},
  author={Panayotov, V. and Chen, G. and others},
  booktitle= "Proc. ICASSP",
  pages={5206--5210},
  year={2015},
}

@misc{kaggleLS,
  author = {{LibriSpeech Clean}}, year = "", 
  note = {\url{https://www.kaggle.com/datasets/victorling/librispeech-clean}},
  url = {https://www.kaggle.com/datasets/victorling/librispeech-clean}}

@misc{kaggleAS,
  author = {{ASVspoof 2019}}, year = "",
  note = {\url{https://www.kaggle.com/datasets/awsaf49/asvpoof-2019-dataset}},
  url = {https://www.kaggle.com/datasets/awsaf49/asvpoof-2019-dataset}}

@article{pot,
  author  = {Flamary, R. and Courty, N. and others},
  title   = {{POT}: Python Optimal Transport},
  journal = {Journal of Machine Learning Research},
  year    = {2021},
  volume  = {22},
  number  = {78},
  pages   = {1-8},
}

@inproceedings{fadtk,
  title = {Adapting {Fr\'echet} Audio Distance for Generative Music Evaluation},
  author = {Gui, A. and Gamper, H. and Braun S. and Emmanouilidou, D.},
  booktitle = "Proc. ICASSP",
  year = {2024},
}

@inproceedings{vgg,
title	= {{CNN} Architectures for Large-Scale Audio Classification},
author	= {Hershey, Sh. and Chaudhuri, S. and others},
year	= {2017},
booktitle	= "Proc. ICASSP"
}

@inproceedings{szegedy2014intriguing,
  title     = {Intriguing Properties of Neural Networks},
  author    = {Szegedy, Christian and Zaremba, Wojciech and Sutskever, Ilya and others},
  booktitle = {International Conference on Learning Representations (ICLR)},
  year      = {2014},
  url       = {https://arxiv.org/abs/1312.6199}
}

@inproceedings{goodfellow2015explaining,
  title     = {Explaining and Harnessing Adversarial Examples},
  author    = {Goodfellow, I. and Shlens, J. and Szegedy, Ch.},
  booktitle = {International Conference on Learning Representations (ICLR)},
  year      = {2015},
  url       = {https://arxiv.org/abs/1412.6572}
}

@misc{papernot2016transferability,
  title   = {Transferability in Machine Learning: From Phenomena to Black-Box Attacks Using Adversarial Samples},
  author  = {Papernot, N. and McDaniel, P. and Goodfellow, I.},
  eprint = {1605.07277},
  year    = {2016},
  note = {arXiv 1605.07277},
  archivePrefix={arXiv},
  primaryClass={cs.CR},
  url     = {https://arxiv.org/abs/1605.07277}
}

@misc{alzantot2018did,
  title   = {Did You Hear That? Adversarial Examples Against Automatic Speech Recognition},
  author  = {Alzantot, M. and Balaji, B. and Srivastava, M. and others},
  eprint = {1801.00554},
  year    = {2018},
  archivePrefix={arXiv},
  note = {arXiv 1801.00554},
  primaryClass={cs.CL},
  url     = {https://arxiv.org/abs/1801.00554}
}

@article{asv,
  title   = {ASVspoof 2021: Towards Spoofed and Deepfake Speech Detection in the Wild},
  author  = {Liu, X. and Wang, X. and  Sahidullah, M. and others},
  journal = {IEEE/ACM Trans. Audio, Speech, Lang. Process.},
  year    = {2023},
  volume= "31",
  pages="2507--2522",
}

@inproceedings{desplanques2020ecapa,
  title={{ECAPA-TDNN: Emphasized} Channel Attention, Propagation and Aggregation in {TDNN} Based Speaker Verification},
  author={Desplanques, Brecht and Thienpondt, Jenthe and Demuynck, Kris},
  booktitle={Proc. Interspeech},
  pages={3830--3834},
  year={2020},
  doi={10.21437/Interspeech.2020-2650}
}
}

\end{document}